# Crowdsourcing public attitudes toward local services through the lens of Google Maps reviews: An urban density-based perspective

Lingyao Li[1*], Songhua Hu[2], Atiyya Shaw[3], Libby Hemphill[1]

[1] School of Information, University of Michigan, Ann Arbor, Michigan
[2] Department of Urban Studies and Planning, Senseable City Lab, Massachusetts Institute of Technology, Cambridge, Massachusetts
[3] Department of Civil and Environmental Engineering, University of Michigan, Ann Arbor, Michigan

*Corresponding author: lingyaol@umich.edu

**Abstract** Understanding how urban density impact public perceptions of urban services is important for informing livable, accessible, and equitable urban planning. Conventional methods such as surveys are limited by their sampling scope, time efficiency, and expense. On the other hand, crowdsourcing through online platforms presents an opportunity for decision-makers to tap into a user-generated source of information that is widely available and cost-effective. To demonstrate such potential, this study uses Google Maps reviews for 23,906 points of interests (POIs) in Atlanta, Georgia. Next, we use the Bidirectional Encoder Representations from Transformers (BERT) model to classify reviewers' attitudes toward urban density and the Robustly Optimized BERT approach (RoBERTa) to compute the reviews' sentiment. Finally, a partial least squares (PLS) regression is fitted to examine the relationships between average sentiment and socio-spatial factors. The findings reveal areas in Atlanta with predominantly negative sentiments toward urban density and highlight the variation in sentiment distribution across different POIs. Further, the regression analysis reveals that minority and low-income communities often express more negative sentiments, and higher land use density exacerbates such negativity. This study introduces a novel data source and methodological framework that can be easily adapted to different regions, offering useful insights into public sentiment toward the built environment and shedding light on how planning policies can be designed to handle related challenges.

# 1. Introduction

The built environment plays an important role in shaping cities' functionality, sustainability, and socioeconomic systems and interactions. More specifically, the built environment is the setting for activities and movement and as a result, impacts accessibility and the quality of services provided (Claassens et al., 2020; McFarlane, 2016). Understanding public opinion about the urban environment is important as it sheds light on citizens' perspectives and needs (L. Yang et al., 2022). In many instances, aligning urban planning initiatives with communities' desires and needs is critical to their success, not to mention a core (and in some cases, mandated) aspect of the planning process (W. Li et al., 2020; Olsson, 2008; Wilson et al., 2019). When citizens perceive urban plans as beneficial in improving their overall quality of life such as through improved access to amenities and/or broadening their environmental experience, they are more inclined to support, utilize, and embrace these initiatives (Kelly & Swindell, 2002; Ochoa Rico et al., 2022). Urban planners and policymakers are therefore key users and stakeholders in developing research and tools that can facilitate the collection and representation of public sentiment and needs (Nigussie et al., 2023).

In addition, understanding the relationships between public attitudes toward the built environment and socio-spatial factors is crucial. It can help reveal the underlying factors that contribute to these attitudes, thereby promoting a more nuanced and comprehensive perspective of living environment preferences (Al Haddad et al., 2020; Salvati et al., 2019; Tu et al., 2020). It also helps to show areas with certain sociodemographic groups that exhibit distinctive perceptions or use patterns in specific activities (Momeni & Antipova, 2022). These insights – either positive or negative – are important to inform rational and tailored policymaking during planning processes; for example, policymakers can prioritize strategies focusing on improving areas with more negative attitudes.

Conventional methods like surveys are long-standing tools used to measure and understand public attitudes toward urban issues (Lu et al., 2017; Paköz & Işık, 2022; Schmidt-Thomé et al., 2013). With the structured questions and quantitative nature, surveys have been effectively applied to gain public opinions toward urban planning initiatives (Lu et al., 2017; Paköz & Işık, 2022). Nevertheless, facilitating more effective policies in urban planning necessitates more comprehensive community engagement. The limitations of conventional surveys could lie in their limited geographic scope further hindered by a range of biases, which exacerbate a lack of representativeness for planning efforts that target a broader population. While large-scale surveys could offer broader insights, they come with trade-offs of being time-consuming and costly (Jones et al., 2013).

Meanwhile, social networks and online review platforms have become increasingly popular for people to communicate opinions and feelings with and to each other (Heinonen, 2011). These platforms provide a virtual channel that allows for information dissemination and collection in ways that are faster, more general, and less constrained by social and geographical restrictions (Bendimerad et al., 2019; Cheung & Thadani, 2010). As a result, crowdsourcing through social networks and online review platforms presents an attractive source of information for decision-makers to understand public attitudes toward urban environments and systems (Song et al., 2020; Williams et al., 2019). Compared to surveys, crowdsourcing through online platforms offers several advantages, including large-scale availability and cost-effectiveness (T. Hu et al., 2017; Wu et al., 2020). It also provides an advantage in passive data collection that does not require users to actively participate in a survey (L. Li et al., 2022, 2023).

Our study seeks to explore the potential of crowdsourcing through online reviews (specifically, Google Map reviews in this paper) for obtaining public attitudes toward the built environment, particularly focusing on urban density. To accomplish this, we select Atlanta, Georgia, in the United States, as the case study and collect reviews for the points of interest (POIs). We then employed multiple natural language processing (NLP) techniques, including Bidirectional Encoder Representations from Transformers (BERT) text classification and sentiment analysis, to process the textual information. Our study aims to address two primary research questions:

- **RQ1**: How can Google Maps reviews be used to sense public attitudes related to urban density?
- **RQ2**: Are there significant connections between public attitudes (derived from Google Maps reviews) and nearby socio-spatial factors?

The first question aims to process human-generated content in shedding light on public attitudes toward urban density in Atlanta. These reviews can contribute to a more nuanced understanding of how surrounding urban density and accessibility impact public perceptions of local services, offering detailed insights that traditional sources may overlook. The second question addresses the potential to cross-reference the insights obtained from Google Maps reviews with local sociodemographic data. These patterns can offer an important perspective on how socio-spatial factors could be associated with urban planning and development.

## 2. Literature review

In this literature review, we first discussed varying definitions of urban density, followed by a look at objective and subjective/perceived measures of the built environment and urban density. In line with the data source and methodological approach of this paper, we concluded the review with an exploration of the use of crowdsourcing data to study urban systems.

### 2.1. The definition of urban density and its impacts

The concept of urban density has been widely used in urban-related studies; examples of these include, public health (Carozzi, 2020), built environment (Ouyang et al., 2020), and transportation systems (Basheer et al., 2020) studies. As might be expected for a concept of interest to several domains, its definition and measurement exhibit diverse interpretations across the literature. In an earlier study, Churchman (1999) suggested that the understanding of urban density should incorporate more significant yardsticks beyond the physical elements, including resources, services, and social interactions, among others. Another widely recognized framework that reflects the concept of urban density is the "3D" framework (Cervero & Kockelman, 1997), which highlights the impacts of the built environment on human activities through three main aspects: density, diversity, and design. This framework was further extended into a "5D" framework by adding the dimension of destination accessibility and distance to transit (Ewing & Cervero, 2001, 2010). Since their development, these D-frameworks have been widely applied in assessing and shaping urban design and land planning (Duan et al., 2023; Lu et al., 2017, 2018).

Meanwhile, other studies have emphasized the complexities and/or nuances of measuring urban density itself. For example, Banai and DePriest (2014) suggested that urban density should encompass the entire urban footprint rather than solely focusing on residential areas in urban sprawl studies. Susanti et al. (2016) identified the indicators of residential density that best fit the character of housing in Indonesia from three perspectives, including built environment, unbuilt environment, and transportation network. Dovey and Pafka (2020) noted that urban density constitutes several multifaceted elements, including density that focuses on people and places within walkable distances, functional mix that produces a range of walkable destinations, and access networks that mediate traffic flows between them. Similarly, Angel et al. (2021) decomposed the concept of urban density into a constitution of different components such as floor space occupancy and floor area density.

While the literature provides a thorough discussion of urban density, taken together it affirms that urban density is a complex concept. Notably, its influence on urban livability remains a subject of debate in the scholarly literature (Keil, 2020). In a study of the high-density city of Kolkata in India, Bardhan et al. (2015) suggested that a compact city policy for high-density cities like Kolkata can help enhance the quality of urban life. Through the case study of Istanbul after the COVID-19 pandemic, Paköz and Işık (2022)) revealed a positive impact of high urban density on urban vitality but a negative impact on health. Similarly, Claveria et al. (2023) investigated developing Asian countries, and their estimates showed that higher population density in urban settings led to increased household wealth, improved hygiene, and enhanced access to utilities and private goods while contributing to heightened outdoor pollution. In addition, (Butters et al., 2020) claimed that dense high-rise urban areas, in comparison to low-dense and medium-rise areas, can not easily bring advantages concerning transport or city sustainability. Duranton and Puga (2020) argued that density might be synonymous with crowding and make living and moving in cities more costly.

## 2.2. Subjective/perceived vs. objective measures of the built environment

A majority of studies in the literature (see prior section) have defined or assessed built environment measures, a core measure of which is urban density, using objective measures. However, it is widely acknowledged that differences exist between objectively measured and stated perceptions of the built environment. Objective measures are derived from data collected in the field, systematic statistics, and spatial information. In contrast, subjective measures are typically obtained through field questionnaire surveys, directly capturing individuals' self-reported perceptions and feelings. For example, Kent et al. (2017) surveyed 562 households in Sydney, Australia to compare the effects of objective and perceived urban built environment on citizen happiness. They found that the perceived built environment was more often associated with subjective well-being compared to the objective-built environment. Another study found low agreement between perceived and objective measures of street connectivity and mixed land use (Koohsari et al., 2015). More recently, Guo and He (2021) compared the role of objective and perceived measures of the built environment in affecting adopting dockless bike-sharing as a feeder mode choice of metro commuting. Paköz and Işık (2022) surveyed 337 participants living in Istanbul in 2020 to understand the relationship between urban density, urban vitality, and a healthy environment. Their study revealed a significant difference between the density levels of the districts in terms of the perception of urban vitality and some sub-variables of a healthy environment.

These findings underscore the necessity of developing reliable and cost-effective tools to collect and measure public perceptions of the built environment and urban density. Surveys remain the core tool for obtaining such data (Lu et al., 2017; Paköz & Işık, 2022; Pérez, 2020). For example, a field questionnaire survey using the Likert scale is a common approach to gathering individuals' self-reported perceptions of the environment. However, due to the high labor and monetary costs, most surveys are conducted on a small scale, and cannot cover a broader set of population groups and activity types at a city scale. These limitations directly motivate this study: i.e., can we obtain perceptions of the built environment from online review platform data?

## 2.3. Crowdsourcing through social networks and review platforms to study urban systems

Crowdsourcing can help gather ideas, information, or contributions from a vast pool of individuals, and. is typically classified into two categories: active and passive. In active crowdsourcing, decision-makers play a proactive role, presenting specific policies or social issues and actively seeking relevant information and opinions from the public. In contrast, passive crowdsourcing involves decision-makers gathering and analyzing content related to a particular topic or public policy that the public have already generated across various sources (Bubalo et al., 2019; Loukis & Charalabidis, 2015). In this study, we utilize passive crowdsourcing, i.e., obtaining opinions about urban density from Google Maps reviews.

In contrast to traditional surveys, crowdsourcing through social networks and online review platforms offers several advantages, including large-scale availability, cost-effectiveness (T. Hu et al., 2017; Wu et al., 2020), and the passive collection of information (L. Li et al., 2022, 2023). Previous research has showcased the remarkable potential of leveraging crowdsourcing via social networks to decipher public sentiments and attitudes across various domains such as business management, healthcare, and even beyond, demonstrating its versatility and applicability in diverse fields (S. Hu, Wang, et al., 2023; L. Li et al., 2020; Paköz & Işık, 2022; Williams et al., 2019).

In the domain of urban-related studies, only a few studies have explored its value. For example, Misra et al. (2014) discussed the potential of crowdsourcing for transportation planning and operations through multiple cases, including bike routes and transit quality. Williams et al. (2019) collected data from several Chinese social media platforms (e.g., Baidu Map) and developed a computational model based on residential visited locations to identify Ghost cities in China – areas marked by housing vacancies resulting from excessive development relative to their small community size. Chen et al. (2019) collected Facebook data and used the residents' visits to POIs to investigate the urban vibrancy in Hong Kong. Their study sought to understand spatial structures and leverage social media insights to enrich urban planning through a nuanced understanding of human activities. Similarly to this study, Chuang et al. (2022) utilized more than 20 million geolocated tweets from July 2012 to October 2016 regarding park visits in Singapore. They further applied statistical methods to examine the influence of park facilities on both spatial density and diversity and revealed that high surrounding population density and family-oriented facilities (e.g., playgrounds) could result in higher spatial density of visitors. More recently, S. Hu et al. (2023) used location-based service data to measure the volume of census block group-level human activities across the whole United States. They also compared a set of machine learning methods to interpret the relationships between human activities and socio-spatial factors.

These studies showcase the potential of using crowdsourcing data to facilitate the urban planning process. However, two questions remain. First, most of these investigations rely on the number of visits from social media posts to unveil urban vibrancy, but few have directly examined the textual content shared by social media users, which could provide direct insights into people's sentiments and opinions. Second, there's a noticeable lack of research specifically focusing on public perception of urban density, possibly due to its complex nature that demands nuanced expertise to decode people's opinions beyond mere visitation frequency. To address these challenges and explore the potential of crowdsourcing, we proposed two research questions outlined in the introduction.

# 3. Data and methods

**Figure 1** presents the research framework applied in this study. The research scheme began with data collection and preparation; this involved collecting Google Maps reviews based on the POIs in Atlanta, Georgia, United States. The resulting dataset contained 23,906 POIs with reviews (**Section 3.1**). Then, we applied an iterative word screening process to identify word filters that could help identify comments that possibly relate to residents' attitudes toward urban density-related features and characteristics (**Section 3.2**). Next, we developed text classification models with word embedding and classifiers to identify comments that imply "actual" urban density-related features (**Section 3.3**). After that, we applied a sentiment tool to calculate the fine-grained sentiment from reviews based on sentence units (**Section 3.4**). With sentiment calculated, we employed a partial least squares (PLS) regression to explore the relationships between average sentiment and socio-spatial factors (**Section 3.5**).

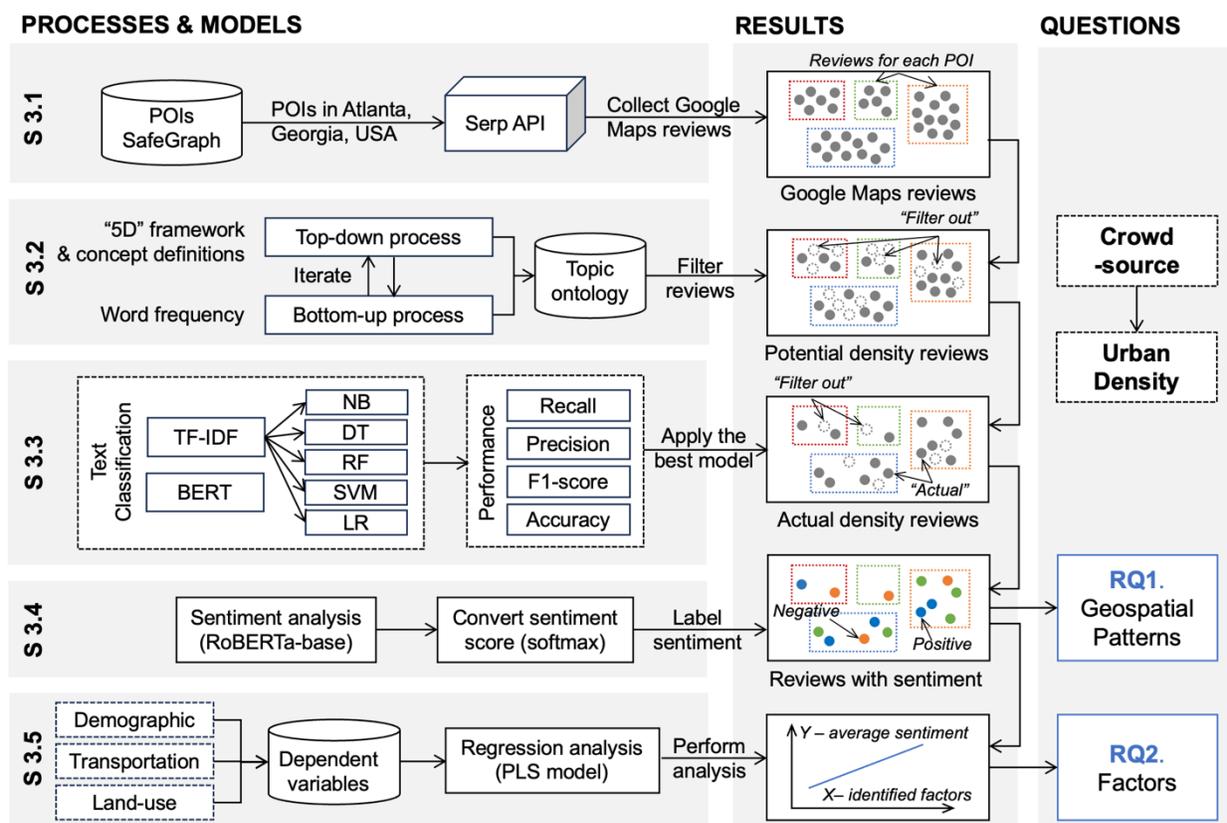

**Figure 1**. Research framework to implement the project (with corresponding sections on the left hand).

## 3.1. Data collection

We selected Atlanta, Georgia, in the United States as the case study for multiple reasons. Firstly, Atlanta is known for its diverse urban landscape, with various neighborhoods, commercial districts, and suburban areas (X. Yang & Lo, 2003). Secondly, Atlanta has risen to become a major transportation hub and one of the fastest-growing metropolitan areas in the United States (Z. Hu & Lo, 2007). Studying its urban density can provide insights into managing urban sprawl and the impact of rapid growth on infrastructure and community dynamics. Thirdly, the city's socioeconomic diversity and historical context (Ambinakudige et al., 2017) provide rich data for understanding the relationship between local built-environment factors and social dynamics.

We sourced the list of Points of Interest (POIs) in Atlanta through Safegraph, an online POI database (SafeGraph, 2023). Based on the addresses of POIs, we used a tool called SerpAPI (SerpAPI, 2023) to search and collect Google Maps reviews associated with each POI. Google Maps is a web mapping platform and consumer application developed by Google. On Google Maps, people can freely rate a place and share their experiences, feelings, and suggestions about a business site, such as a restaurant, scenic spot, commercial district, or airport. We opted for Google Maps data for two primary reasons. Firstly, it has shown a substantial surge in reviews since 2015, surpassing other platforms like Yelp or TripAdvisor (Munawir et al., 2019). Secondly, compared to social media data, Google Maps reviews predominantly reflect customer experiences with businesses, making it a reliable source for our crowdsourcing implementation (Lee & Yu, 2018).

Using SerpAPI, each downloaded Google Maps review contained information about a reviewer's username, rating, number of likes, review text, and posted images. The Google Maps rating system employs a scale of one to five stars with one star for poor and five stars for excellent. To ensure inclusivity, non-English reviews were translated by Google into English during the collection phase for text analysis. Our data collection process encompassed 23,906 POIs in Atlanta while 12,839 POIs from SafeGraph couldn't be collected with the SerpAPI (possibly due to their address change or name change) or were not associated with any reviews. Reviews for each POI were saved in separate JSON files, which were used for subsequent analysis.

## 3.2. Topic ontology and development

We implemented an iterative top-down and bottom-up process to establish the topic ontology, as presented in the box of "S 3.2" in **Figure 1**. This process aimed to compile a set of words to filter in reviews that possibly imply urban density. For the top-down process, we aimed to identify the overarching aspects indicative of urban density. To achieve this, we referenced the well-regarded "5D" framework developed by , where "density" measures the concentration of people, services, and jobs, "diversity" denotes the mixture of land use functions, "design" indicates the quality of surrounding built environment, "destination accessibility" captures the potential of spatial interactions with destination opportunities, and "distance to transit" represents the distance to nearest transit stations. In addition, we drew insights from prior urban density-related studies (Banai & DePriest, 2014; Dovey & Pafka, 2020), with a particular emphasis on three aspects within the 5D framework — density, destination accessibility, and distance to transit — as key metrics for understanding urban density. We also noticed that urban density-related terms like "parking," "walkable," and "within walking distance," which often appeared in the reviews, also implied people's attitudes toward the destination accessibility and distance to transit. Guided by this prior research, we conducted a brainstorming session to generate relevant words indicative of urban density-related features.

For the bottom-up process, we manually screened the 10,000 most frequently occurring words in the dataset. This approach can help ensure the coverage of most keywords (excluding common and ambiguous words). To perform this manual screening, all the terms were first ranked based on their frequencies in the reviews. Since words with higher frequency reflected what residents cared about, the screening process allowed us to gain insights into the determination of words indicative of urban density-related features. This method served as a bottom-up process, offering a broad understanding of the word landscape associated with urban density in Google Maps reviews.

In addition, we noted that individuals often did not think or use words that directly correspond to particular aspects (e.g., dimensions in the 5D framework) in the top-down process, which were typically collected from objective measures/sources. Therefore, we combined the terms identified in the top-down and bottom-up processes into one category, aiming to build a database of reviews that used objective built-environment related measures and characteristics as well as those that used colloquial terms that people used to express their perceptions, attitudes, and emotions toward their experiences with the built environment.

Next, words in the topic ontology were reviewed by each author to make alterations and to ensure the coverage of related terms. The ontology was finalized with two iterative loops (i.e., each loop consists of a top-down and bottom-up process) of development and alterations. The final lexicon-based ontology is presented in **Table 1**. We used these terms to filter the dataset and found that 11,158 of the collected POIs have potential urban density-related reviews.

**Table 1**. Topic ontology for indication of potential urban density-related reviews.

| Collected terms |
| --- |
| 'area', 'access', 'accessibility', 'accessible', 'adjacent', 'bustling', 'closer', 'close to', 'congest', 'congestion', 'convenience', 'convenient', 'dense', 'density', 'distance', 'district', 'easy to find', 'far from', 'location', 'navigate', 'navigation', 'near', 'nearby', 'neighbor', 'open space', 'outdoor', 'packed', 'parking', 'pedestrian', 'populated', 'proximity', 'reachable', 'region', 'right next to', 'surrounded', 'surrounding', 'traffic', 'travel time', 'travel distance,' 'transit', 'transport', 'transportation', 'walkable', 'walkability' |

### 3.3. Text classification

Nevertheless, these descriptive words in the topic ontology could appear in reviews that were not informative about people's feelings about urban density-related features. Examples of such reviews are attached in **Appendix A.1**. These were considered "False" urban density comments (i.e., reviews containing typically descriptive words but not informative of urban density). For example, the term "crowded" could refer to the dense population and traffic flow outside a store or the high foot traffic inside the store, while only the former was related to the urban density that this study focused on. Therefore, we built text classification models to identify whether a comment is associated with urban density attitudes (see the box of "S 3.3" in **Figure 1**), aiming to filter out those "False" reviews.

To establish the training dataset, we randomly selected 1,000 unique reviews from the filtered dataset in **Section 3.2** and manually classified them into either "True" or "False" attitudes toward urban density. Two authors with expertise in urban planning and transportation labeled these reviews. As a result, 376 and 624 of the 1,000 reviews were judged to describe "True" (e.g., sampled reviews in **Table 2**) and "False" urban density attitudes (e.g., sampled reviews in **Table A.1**), respectively. It should be noted that the manual labeling of these samples was not based on objective or observed estimation but on subjective judgments based on the descriptions of urban density discussed in **Section 2.1**. Following this approach, we randomly selected another 250 unique samples (different from training samples) as the testing dataset, in which 95 and 155 were identified as "True" and "False" samples, respectively.

Next, we applied NLP and machine learning techniques to build text classification models. We mainly applied two NLP techniques of word embeddings, namely Term Frequency-Inverse Document Frequency (TF-IDF) (Ramos, 2003) and Bidirectional Encoder Representations from Transformers (BERT) (Devlin et al., 2018). These techniques of word embeddings can help convert each review into a vector or a matrix of numbers that can be processed by the machine. The candidate machine learning classifiers used in conjunction with the TF-IDF vectorization technique included Decision Tree (DT), Random Forest (RF), Naïve Bayes (NB), Support Vector Machine (SVM), and Logistic Regression (LR). The BERT model we applied was the BERT sequence model for text classification, where the classifier was built-in. This resulted in six candidate model combinations: TF-IDF + DT, TF-IDF + RF, TF-IDF + NB, TF-IDF + SVM, TF-IDF + LR, and BERT.

For each of the candidate models, we performed hyperparameter tuning within the training partition using K-fold cross-validation with K=5 in this project. That is, in each loop during the training process, the model was trained using four subsets and validated against the remaining subset. Through the 5-fold cross-validation, we tested different hyperparameters in each candidate model using the grid search, as presented in **Appendix A.2**. Then, we selected the best hyperparameter for each model based on the average performance (i.e., accuracy) over the testing values returned by 5-fold cross-validation.

The performance of the six candidate models with tuned hyperparameters was assessed using four metrics: Precision, Recall, F1-score, and Accuracy. Accuracy is the fraction of correctly classified cases over all cases. Precision measures the fraction of true positive cases over the retrieved cases a model predicts. Recall is the fraction of true positive cases over all the relevant cases that are identified. F1-Score combines precision and recall and reflects the model's prediction capability. The performance of each model is presented in **Figure 2**.

The selection of the "best-performing" model was mainly based on the testing accuracy and F1-score (which combines Precision and Recall). **Figures 2(c)** and **2(d)** show that BERT obtains the highest testing accuracy of 83.6% and a more balanced F1-score between the "True" and "False" classes compared to other models. We therefore selected the BERT model as the "best-performing" model and applied it to the whole dataset. After that, we only kept the "True" reviews for sentiment analysis, as those reflected residents' attitudes toward urban density. As a result, only 8,993 POIs in the dataset contained "True" reviews in terms of their representation of urban density attitudes.

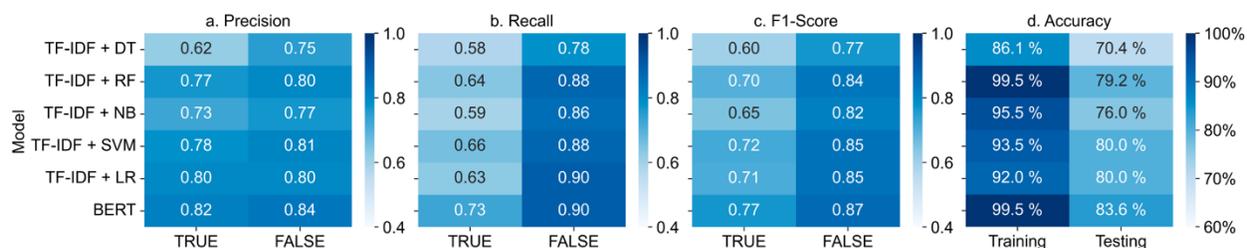

**Figure 2**. Classification performance of candidate models: (a) Precision, (b) Recall, (c) F1-score, and (d) Training and testing accuracy.

In addition, we observed that certain POIs only have limited urban density-related reviews. Incorporating these POIs with minimal reviews could disproportionately influence the sentiment outcomes, as merely one or two reviews could produce a significant impact. To counteract this bias, we selectively opted to retain POIs with 10 or more reviews. This curation could help ensure more representative data, which resulted in a refined dataset comprising 1,184 POIs for the following sentiment and regression analysis.

### 3.4. Sentiment classification

We selected the sentence as the unit for sentiment analysis since some reviews can be long and include aspects not indicative of urban density. Such fine-grained sentiment classification is advantageous when a person expresses opposite sentiments regarding different aspects of a review. Therefore, we segmented each review into multiple parts by the sentence endings, namely a period (.), a semicolon (;), a question mark (?), an exclamation point (!), or an ellipsis (…). The sentences related to the urban density were extracted based on terms in **Table 1**, and examples of these sentences are presented in the second column in **Table 2**.

Next, we applied a RoBERTa-base sentiment tool to compute the sentiment, as presented in the box of "S 3.4" in **Figure 1**. RoBERTa is a robustly optimized BERT pre-training approach, which was built on the transformer architecture with an attention mechanism to generate contextualized representation. The RoBERTa-base sentiment model was fine-tuned for sentiment analysis with the TweetEval benchmark, which specifically focuses on analyzing tweet sentiment. Barbieri et al. (2020) added a dense layer to reduce the dimensions of RoBERTa's last layer to the number of labels in the classification task to prepare the model for sentiment classification. This model has demonstrated its superior performance over FastText and SVM-based models with n-gram features (Barbieri et al., 2020). Loureiro et al. (2022) further updated the model by training it on a larger corpus of tweets, based on 123.86 million tweets extracted until the end of 2021, compared to its predecessor's 58 million tweets.

By leveraging the latest model, we were able to analyze the sentiment of reviews that mentioned urban density-related views. It is important to recognize that the sentiments expressed in these reviews primarily convey people's perceptions of density-related features rather than indicators of their support for high or low-density urban features. These reviews only reflect individuals' subjective feelings and experiences when interacting with the built environment. Examples and their resulting sentiment classifications are presented in **Table 2**. The sentiment classification follows the highest score returned by the softmax layer in the RoBERTa model. Examples of sentiment classification are presented in **Table 2**.

**Table 2**. Examples of sentiment analysis from Google Maps reviews.

| Reviews | Urban density-related reviews | Sentiment score | Sentiment classification |
|---|---|---|---|
| I looooved Alexus! She definitely took care of me and my home girl. Parking: parking was reasonable. | Parking: parking was reasonable. | 'Negative': 0.0090, 'Neutral': 0.3064, 'Positive': 0.6846 | Positive |
| First rate top notch food and service. If you like ice you get a whole lot in your margaritas! Parking can be a pain at times. Worth the wait, most of the time. | Parking can be a pain at times. | 'Negative': 0.8505, 'Neutral': 0.1366, 'Positive': 0.0129 | Negative |
| Beautiful hotel. Very affordable and close to many downtown attractions. Within walking distance of Centennial Olynpic Park [sic]. | Very affordable and close to many downtown attractions. Within walking distance of Centennial Olynpic Park [sic]. | 'Negative': 0.0023, 'Neutral': 0.1109, 'Positive': 0.8868 | Positive |
| Beautiful events facility and a caveat to have ASW Distillery on the property. I've attended multiple events in the past 2 years and all have been outstanding. Only challenge traffic is difficult coming from any direction. | Only challenge traffic is difficult coming from any direction. | 'Negative': 0.6838, 'Neutral': 0.2966, 'Positive': 0.0196 | Negative |
| It's a fairly recently updated hotel on the west side of Atlanta that provides good value. Convenient to Six Flags although that wasn't the purpose of my trip. Very clean and a good value. Just outside perimeter but 15 minutes from downtown. | Convenient to six flags although that wasn't the purpose of my trip. | 'Negative': 0.0705, 'Neutral': 0.7212, 'Positive': 0.2083 | Neutral |

## 3.5. Partial least squares (PLS) regression

With sentiment calculated, we employed PLS regression at the census block group level to investigate the relationships between average sentiment and socio-spatial factors, as shown in the box of "S 3.5" in **Figure 1**. The summary of variables are reported in **Table 3**. Among socio-spatial factors, we obtained socioeconomic and demographic characteristics from the 2022 American Community Survey (ACS) 5-year estimates. Additionally, transportation and land use features were sourced from the Georgia Association of Regional Commissions and the Atlanta Department of City Planning. All socio-spatial factors were aggregated at the census block group (CBG) level, as this provides the finest resolution available. To maintain consistency, we calculated the average POI-level sentiment at the CBG level, weighting it by the number of reviews. Only CBGs with a total review count exceeding 10 were retained, resulting in a dataset of 223 CBGs used for PLS fitting.

PLS is particularly used to address multicollinearity among independent variables. In our case, several variables, including *college degree*, *median income*, *white*, and *African American*, exhibited high variance inflation factors (VIF) exceeding 5, indicating significant multicollinearity. While dropping collinear variables is a common approach to alleviate multicollinearity, it leads to the loss of some important variables of interest.

Through PLS regression, rather than directly regressing the dependent variable on independent variables, we constructed a linear regression by decomposing both the dependent variable and independent variables into orthogonal scores and loadings. The regression coefficients were then estimated using these scores, and a final step was involved to convert them back to the original variable space (refer to **Appendix A.3** for details) (De Jong, 1993). It is worth noting that traditional PLS does not offer p-values for estimated coefficients. In this study, we computed p-values for regression coefficients based on Jack-Knifing resampling with 10-fold cross-validation by comparing the perturbed model coefficient estimates from cross-validation with the estimates from the full model (Martens & Martens, 2000).

**Table 3.** Summary of variables in PLS regression.

|  | Description | Mean | St.d. |
|---|---|---|---|
| Dependent variable | | | |
| Average Sentiment | CBG-level average sentiment extracted from Google Reviews, from -1 (most negative) to 1 (most positive). | 0.381 | 0.246 |
| Independent variables | | | |
| *Socioeconomic and demographic characteristics* | | | |
| Pct. College Degree | Percentage of people with education attainment equal to/higher than college | 52.540 | 26.819 |
| Median Income | The median household income, in $10^3$/household | 72.021 | 44.619 |
| Pct. White | Percentage of people identifying as Non-Hispanic White | 45.110 | 32.321 |
| Pct. African American | Percentage of people identifying as African American | 44.897 | 35.500 |
| Pct. Hispanic | Percentage of people identifying as Hispanics/Latino | 6.577 | 9.984 |
| Pct. Asian | Percentage of people identifying as Asians | 4.998 | 7.382 |
| Pct. Age 18-44 | Percentage of residents between 18 and 44 years | 46.929 | 16.724 |
| Pct. Age 45-64 | Percentage of residents between 45 and 64 years | 23.210 | 8.944 |
| Pct. Age over 65 | Percentage of residents 65 years and over | 12.030 | 9.056 |
| Pct. Male | Percentage of residents identifying as male | 48.191 | 7.875 |
| Population Density | Population density, in $10^3$ persons/sq. mile | 5.476 | 5.912 |
| *Transportation* | | | |
| Bus Stop Density | Bus stop density, in count/sq. mile | 32.322 | 29.766 |
| Metro Station Density | Metro station density, in count/sq. mile | 0.330 | 1.600 |
| Primary Road Density | Primary road density, in mile/sq. mile | 4.693 | 4.722 |
| Secondary Road Density | Secondary road density, in mile/sq. mile | 14.107 | 5.235 |
| Minor Road Density | Minor road density, in mile/sq. mile | 12.226 | 7.947 |
| *Land use* | | | |
| Pct. Industrial Land | Percentage of industrial land use | 0.591 | 3.573 |
| Pct. Institutional Land | Percentage of institutional land use | 4.650 | 9.962 |
| Pct. Utilities Land | Percentage of transportation, communication, and utilities land use | 5.020 | 7.128 |
| Pct. Commercial Land | Percentage of commercial land use | 18.522 | 18.382 |
| Pct. Residential Land | Percentage of residential land use | 53.432 | 24.565 |
| LUM [a] | Land use mixed index | 0.610 | 0.183 |

a. LUM (land-use mixture entropy) was calculated to represent the degree of mixed land use, varying from 0 (homogeneous) to 1 (heterogeneous) (Manaugh & Kreider, 2013):

$$LUM = \begin{cases} \frac{-1}{ln(N)}\sum_{i=1}^{N} p_i ln(p_i) & N > 1 \\ 0 & N = 1 \end{cases} \quad \text{Eq. 1}$$

where $N$ is the number of land-use types in the CBG and is the percentage of land in type $i$.

# 4. Results

The results section includes three subsections. **Section 4.1** addresses the first research question to demonstrate the potential of Google Maps reviews to capture public attitudes toward urban density-related features. **Section 4.2** explores the second research question and aims to illustrate the connections between public attitudes and related socio-spatial factors. In **Section 4.1**, we first presented the distribution of POIs for analysis along with the geospatial patterns depicting sentiment distribution. Then, we reported the results of statistical testing to discern subtle variations in these sentiments toward different POI types. After that, we presented the results of semantic analysis to comprehend public concerns regarding divergent sentiments across POI types. In **Section 4.2**, we presented results from the PLS regression model to show factors that are associated with sentiments toward urban density-related features.

We defined the POI types based on their North American Industry Classification System (NAICS) code. **Figure 3(a)** shows the distribution of POI types based on these 1,184 POIs. Most of these reviews were from four categories, including Accommodation and Food (NAICS code 72) (e.g., restaurants, hotels, and coffee shops), Real Estate and Rental and Leasing (NAICS code 53) (e.g., apartments), Retail Trade (NAICS code 44~45) (e.g., supermarket, grocery stores, liquor stores), and Arts, Entertainment, and Recreation (NAICS code 71) (e.g., museums, parks, art centers). Further breakdowns for each top NAICS category, represented in **Figure 3(b)** through **Figure 3(e)**, highlight the top three subcategories. Notably, among all subcategories, Restaurants and Eating Places (**Figure 3(b)**) and Lessors of Real Estate (e.g., apartments) (**Figure 3(c)**) emerge as the most frequently reviewed POIs concerning urban density in our analysis.

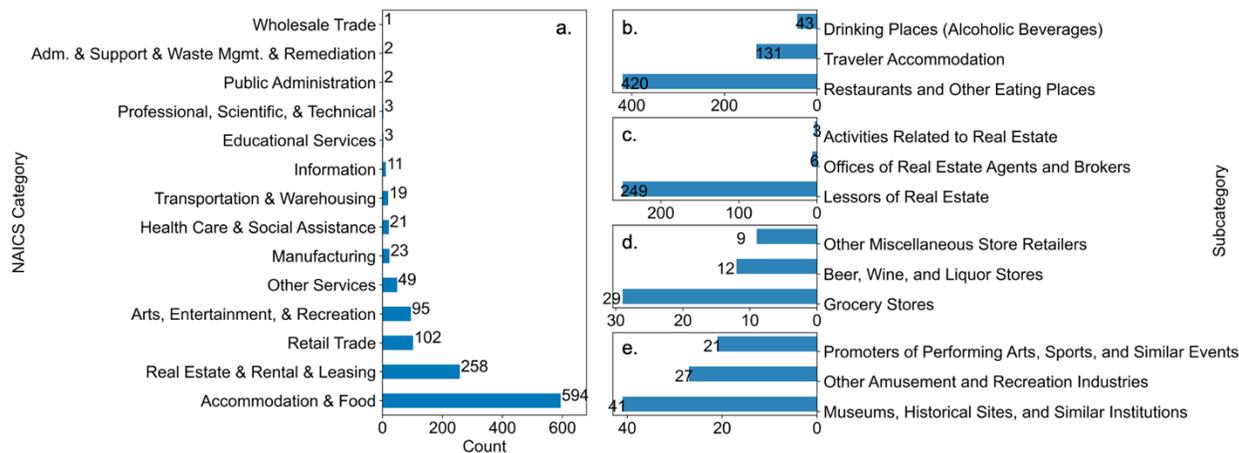

**Figure 3**. The distribution of 1184 POIs for result analysis in the (a) NAICS Category, (b) Category of Accommodation and Food, (c) Category of Real Estate, Rental, and Leasing, (d) Category of Retail Trade, and (e) Category of Arts, Entertainment, and Recreation.

## 4.1. RQ1 – How can Google Maps reviews be used to sense public attitudes related to urban density?

**Figure 4(a)** illustrates the distribution of POIs in Atlanta; each POI is colored according to the average sentiment of reviews about it. Within the pool of 1,184 POIs examined, 1,025, 34, and 125 exhibited positive, neutral, and negative average sentiment, respectively. Overall, the public sentiment toward POIs in Atlanta is positively skewed. In addition, geographic areas in Atlanta show different average sentiments toward urban density as indicated by the orange-colored or blue-colored POIs. For instance, referencing **Figure 4(a)**, Area 1 – Atlanta's Buckhead district, Area 2 – proximity to Freedom Park, and Area 3 – downtown Atlanta, portray more negative sentiments. In contrast, Area 4 – Loring Heights neighborhood,

Area 5 – areas close to Techwood Dr. NW, and Area 6 – downtown Atlanta, exhibited more positive sentiment.

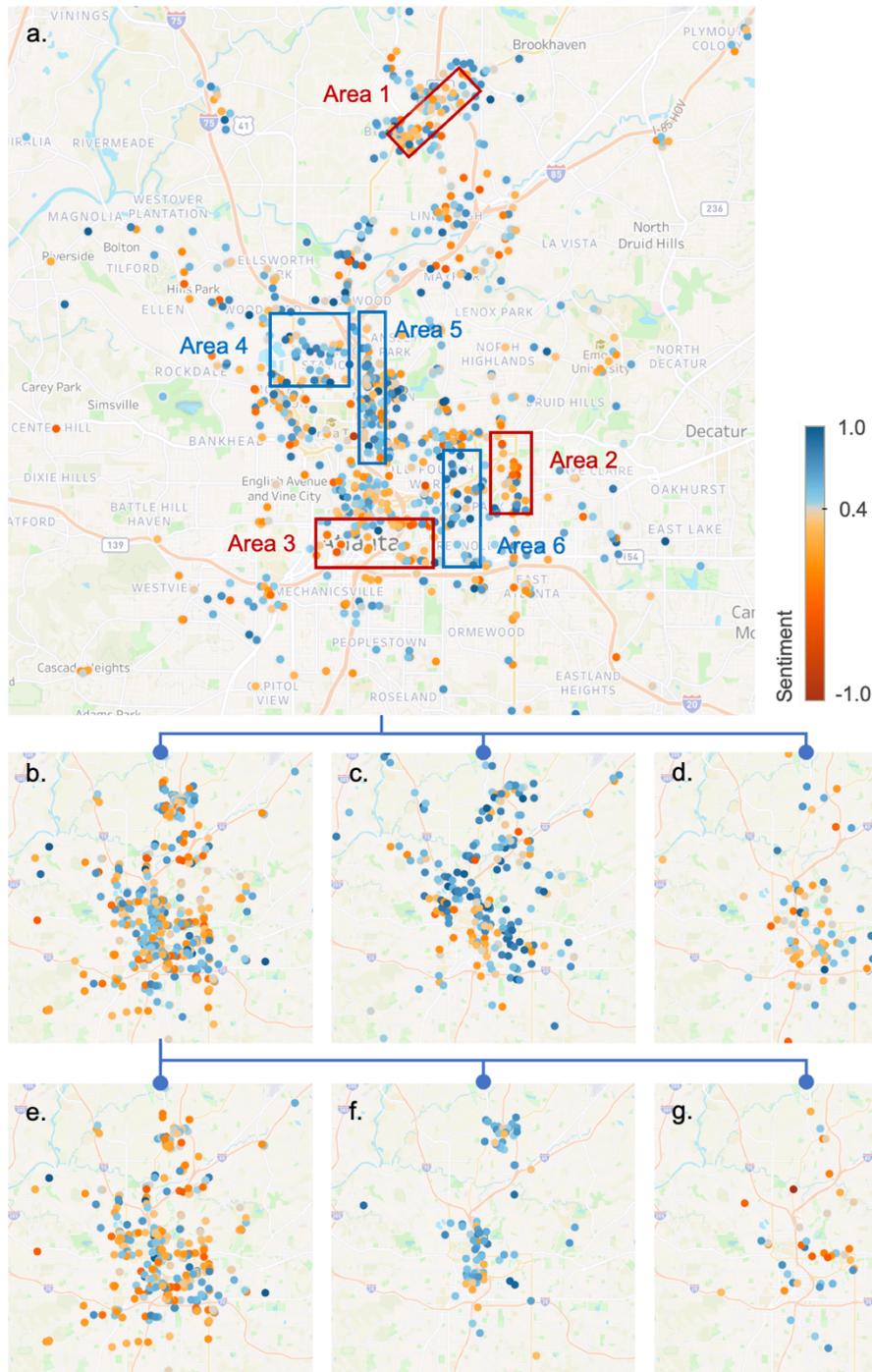

**Figure 4**. Sentiment of POIs in (a) all NAICS categories in Atlanta, (b) Accommodation and Food Services, (c) Real Estate and Rental and Leasing, (d) Retail Trade, (e) Restaurants and Eating Places, (f) Traveler Accommodation, and (g) Drinking Places. Blue indicates more positive sentiment, while orange indicates more negative sentiment; the center of the bar is set at 0.4 as the sentiment distribution is positively skewed.

Upon closer examination of the sentiment distribution for POIs across various NAICS categories, we observed differences in the average sentiment levels. For example, a visual comparison between **Figure 4(c)** and **Figures 4(b)** and **4(d)** illustrates that the average sentiment within the Real Estate and Rental and Leasing category is greater than other categories. As the category of the category of Accommodation and Food Services has the most POIs in our dataset, a further breakdown (**Figure 4(e~g)**) indicates variations in the average sentiment even within the same category. Subsequently, we conducted a Mann–Whitney U test, a.k.a. the Wilcoxon rank-sum test, to examine whether significant differences in average sentiment exist between two POI types without the requirement to meet the normal distribution assumption. The hypothesis testing results are presented in **Figures 5 (a)** and **(b)**.

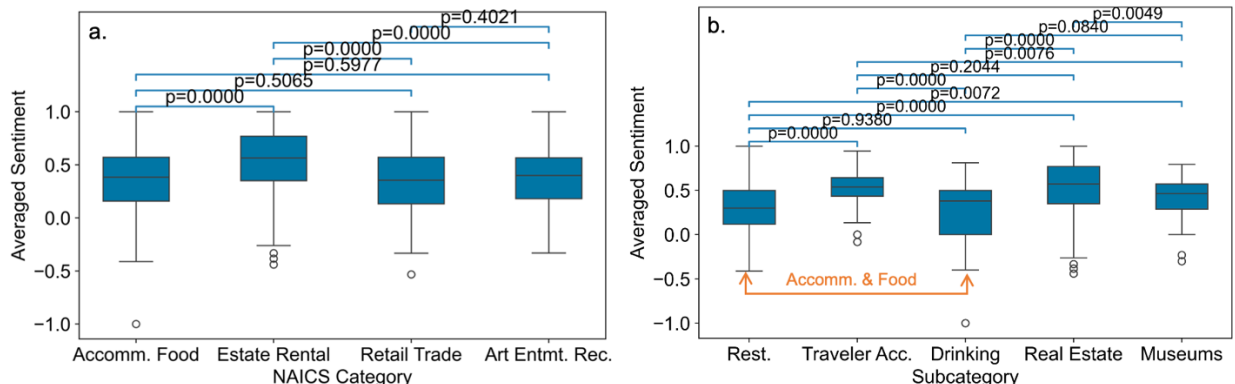

**Figure 5**. Sentiment distribution and Mann-Whitney U test for (a) Top four NACIS categories (according to **Figure 3**) and (b) Major subcategories with number of POIs ≥30.

**Figure 5(a)** illustrates a notable finding – the "Estate Rental" category exhibits a significantly higher average sentiment compared to the other three categories ($p < 0.05$). However, the remaining three categories don't demonstrate a significant difference in sentiment levels. Based on **Figure 5(b)**, there are several observations to highlight. First, within a NAICS category, subcategories can display varying average sentiments. For instance, in the "Accommodation & Food" category, the "Traveler Accommodation" subcategory (e.g., hotels) shows a notably higher average sentiment compared to the other two subcategories – "Restaurants and Eating Places" and "Drinking Places." Second, our analysis reveals significant differences between subcategories across different NAICS categories. For instance, the average sentiment for the "Lessors of Real Estate" subcategory (e.g., apartments) is significantly higher than subcategories found in other NAICS categories, such as Museums and Restaurants. Last, despite the "Traveler Accommodation" and "Lessors of Real Estate" subcategories falling under the NAICS categories, their average sentiments do not display a significant difference. It should be noted that both categories involve living spaces, with "Traveler Accommodation" encompassing hotels and "Lessors of Real Estate" focusing on apartments.

Following the primarily descriptive discussion in **Section 4.1**, we further explored the public perceptions of urban density using a tool called lexical salience-valence analysis (LSVA). This method employs text mining to examine the relationship between words and their sentiments expressed in reviews. Unlike counting the occurrence of words in positive or negative reviews, LSVA offers a more comprehensive understanding by visualizing word frequency across the document corpus and its influence on overall sentiment (please refer to **Appendix A.4** for the details). Using this approach, we graphed the top 30 words with the highest salience (excluding stopwords) along with their valence in **Figure 6(a~d)** across the four NAICS categories. A higher salience denotes a greater prevalence of a word within the dataset, while a higher valence indicates a more positive sentiment garnered from customer reviews.

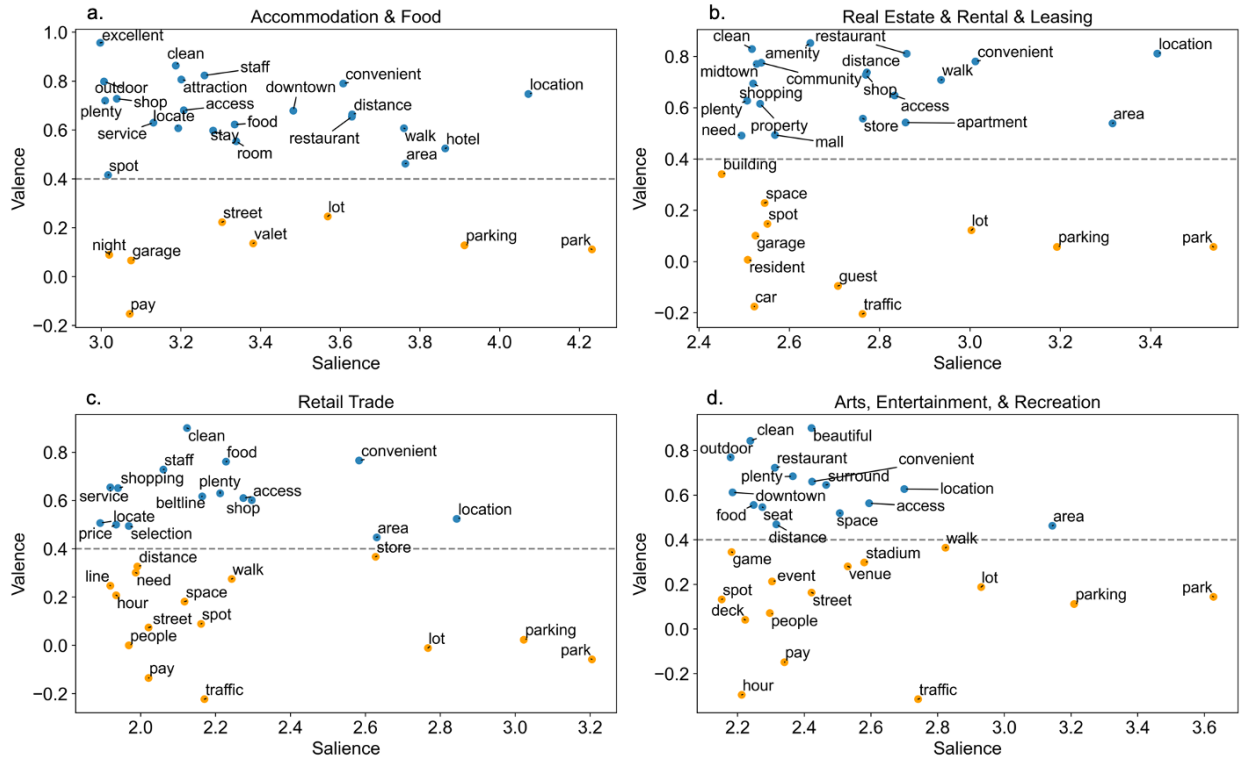

**Figure 6**. The relative importance of words mined from Google Maps reviews for urban density proxied by the LSVA method in the category of (a) Accommodation and Food Services, (b) Real Estate and Rental and Leasing, (c) Retail Trade, and (d) Arts, Entertainment, and Recreation.

**Figure 6** illustrates several consistent patterns. Focusing on salience, it reveals a uniform concern for diverse aspects related to urban density, underscoring elements like parking, access, convenience, location, walking distance, and traffic. In terms of valence, the results indicate that parking-related terms consistently rank lower, whereas access is marked by a higher valence in comparison to other factors. Additionally, aspects such as location and convenience demonstrate higher salience and positive valence values.

Despite the consistent trends, **Figure 6** also highlights distinct concerns within each category. In the Accommodation and Food category encompassing restaurants, hotels, and drinking places, parking emerges as a predominantly negative concern. Within the Real Estate and Rental and Leasing category, primarily consisting of POIs of apartments, community-related facilities such as shopping, stores, amenities, malls, and community spaces get much attention, and they appear generally positive as depicted in **Figure 6(b)**. Regarding the Retail Trade category, primarily comprising stores and supermarkets, walking distance evokes more negative valence compared to the other three categories. In the category Arts, Entertainment, and Recreation which encompasses facilities like gyms and art centers capable of hosting significant events, concerns converge to aspects related to streets, parking, traffic, games, and event management.

### 4.2. RQ2 – Are there any connections between public attitudes and socio-spatial factors?

The results of the PLS regression (1st component) are reported in **Table 4**. Coefficients represent the relationships between average sentiment extracted from Google Maps reviews and the community socio-spatial factors at a CBG level. Since all variables are Z-score standardized, the magnitude of coefficients can be directly compared. Among all independent variables, the percentage of African Americans exhibits the most substantial negative coefficient (-0.073), followed by bus stop density (-0.036) and the LUM (-0.031). On the contrary, the percentage of whites shows the greatest positive coefficient (0.072), followed by the percentage of people with a college (or higher) degree (0.067) and median household income (0.064). These findings align closely with the bivariable Pearson correlation as illustrated in **Figure 7**, where the spatial distribution of sentiment mirrors that of the percentage of whites and inversely relates to the percentage of African Americans. In summary, the PLS regression reveals several notable patterns in public sentiment related to urban density:

1) Racial segregation: Regions with a higher concentration of Whites and Asians tend to show positive sentiment, while those with more African Americans are associated with negative sentiment.
2) Socioeconomic disparity: Regions with higher socioeconomic conditions, including a higher percentage of highly educated individuals and a higher median household income, tend to display positive sentiment.
3) Land and facility development: Regions with denser transportation facilities and mixed land use (e.g., higher LUM, road density, and bus stop density) are more likely to exhibit negative sentiment. Regions with more residential land use tend to show positive sentiment, while those with more industrial land use are more likely to exhibit negative sentiment.

**Table 4.** Results of PLS regression models (Z-score standardized, 1st component).

| Category | Variable | Coeffs | Std. err. | P-value | 2.50% |
|---|---|---|---|---|---|
| Sociodemographics | Pct. College Degree | 0.067 | 0.014 | 0.001*** | 0.036 |
| | Median Income | 0.064 | 0.013 | 0.001*** | 0.034 |
| | Pct. White | 0.072 | 0.013 | 0.000*** | 0.043 |
| | Pct. African American | -0.073 | 0.013 | 0.000*** | -0.104 |
| | Pct. Hispanic | 0.000 | 0.018 | 0.993 | -0.040 |
| | Pct. Asian | 0.019 | 0.007 | 0.019* | 0.004 |
| | Pct. Age 18-44 | -0.001 | 0.020 | 0.958 | -0.047 |
| | Pct. Age 45-64 | -0.005 | 0.014 | 0.719 | -0.038 |
| | Pct. Age over_65 | 0.016 | 0.007 | 0.040* | 0.001 |
| | Pct. Male | 0.012 | 0.013 | 0.359 | -0.017 |
| | Population Density | -0.014 | 0.018 | 0.456 | -0.054 |
| Transportation | Bus Stop Density | -0.036 | 0.022 | 0.003** | -0.053 |
| | Metro Station Density | -0.002 | 0.014 | 0.893 | -0.033 |
| | Primary Road Density | -0.027 | 0.010 | 0.018* | -0.050 |
| | Secondary Road Density | -0.020 | 0.023 | 0.422 | -0.072 |
| | Minor Road Density | -0.004 | 0.010 | 0.724 | -0.026 |
| Land use | Pct. Industrial Land | -0.030 | 0.012 | 0.034* | -0.058 |
| | Pct. Institutional Land | -0.022 | 0.014 | 0.168 | -0.054 |
| | Pct. Utilities Land | 0.004 | 0.014 | 0.779 | -0.028 |
| | Pct. Commercial Land | -0.009 | 0.015 | 0.559 | -0.042 |
| | Pct. Residential Land | 0.024 | 0.009 | 0.031* | 0.003 |
| | LUM | -0.031 | 0.010 | 0.008** | -0.055 |
| Model Goodness-of-fit | | | | | |
| $R^2$ | 0.179 (CV: 0.162) | | | | |
| RMSE | 0.233 (CV: 0.240) | | | | |

**Note:**
a. Significance codes: 0 '***' 0.001 '**' 0.01 '*' 0.05 '.' 0.1 ' ' 1.
b. 0.179 (CV: 0.162) means the $R^2$ for the whole dataset is 0.179 while for 10-fold cross-validation is 0.162.

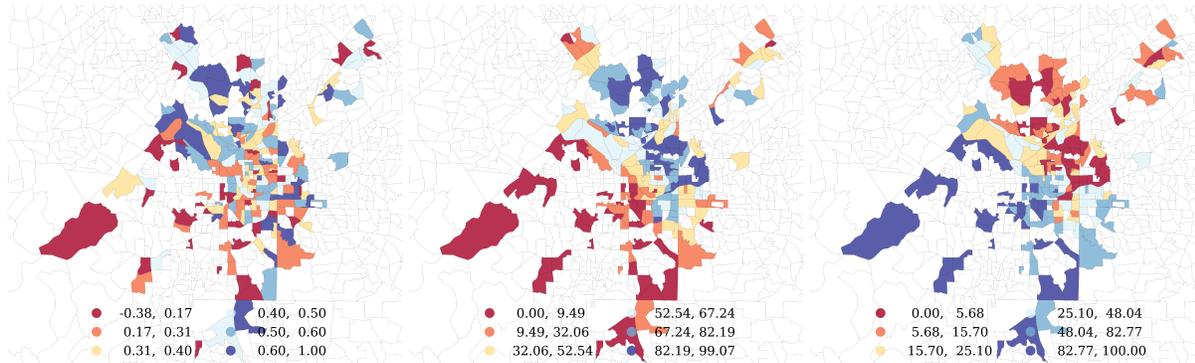

**Figure 7**. Spatial distribution of average sentiment (a) and other factors with the top two highest Pearson's correlation: (b) Pct. of White (0.281), (c) Pct. of African American (-0.283).

## 5. Discussion

Urban density and accessibility are important aspects in shaping the cities' built environment. Crowdsourcing via social media and online review platforms provides an additional resource for mining public perceptions toward urban density. Building on existing research (Chuang et al., 2022; Paköz & Işık, 2022; Song et al., 2020; Williams et al., 2019), our study aims to demonstrate the potential of harnessing crowdsourced data from Google Maps reviews to show how the residents feel about local services that are associated with urban density.

Addressing the first research question, we computed the POI-level sentiment related to urban density from Google Maps reviews and presented a heatmap illustrating its spatial distribution across Atlanta. Our investigation further discerned four predominant POI types in online reviews, unveiling specific concerns regarding urban density within these categories. In response to our second question, we conducted a PLS regression analysis, revealing underlying connections between public sentiment and factors such as racial segregation, socioeconomic disparity, and land and facility development. In sum, this study makes various theoretical and practical contributions to current research by delving into the potential of a novel, crowdsourced data source to depict public attitudes toward urban density, as well as how the extracted attitudes can be used to guide the design and implementation of planning policies.

### 5.1. Theoretical implications

**Propose an iterative approach to developing topic ontology** – The establishment of the topic ontology was guided by a dual process: a top-down approach rooted in domain expertise listed the ontology's subjects, while a bottom-up strategy involved manual screening of the most prevalent terms sourced from Google Maps reviews. This resulting topic ontology could function as a repository for machine learning models to glean insights from extensive collections of online reviews about urban density aspects in built environment.

**Demonstrate the value of crowdsourcing from the lens of online reviews** – Residents in their daily use of city services have direct sentiments and experiences related to urban services. Mining insights from these experiences has become pivotal for urban designers in their quest to enhance urban planning and management (Abdul-Rahman et al., 2021). Leveraging crowdsourcing taps into the potential of "citizens-as-sensors" (Goodchild, 2007) in this context, presenting a distinctive viewpoint on urban density aspects. Compared to conventional survey methodologies, this crowdsourcing approach has several advantages, including data scalability, cost-effectiveness (T. Hu et al., 2017; Wu et al., 2020), and passive data collection without the necessity for active survey participation from users (L. Li et al., 2022, 2023).

## 5.2. Practical implications

**Identify geospatial zones with urban density challenges** – This approach allows urban decision-makers to identify both satisfactory and unsatisfactory zones from public reviews within a city. For instance, our case study reveals several geospatial areas (**Figure 4**) with more positive or negative sentiments in Atlanta, Georgia. In areas with negative sentiment, urban designers can strategize improvements such as optimizing road network designs, expanding parking capacity, enhancing public transit services, reducing land development intensity, and increasing the area of open spaces or green spaces to mitigate the public negative feelings and experiences induced by pro-density development.

**Provide insights into density-related public concerns by POI types** – The comparison analysis across different POI types provides type-specific suggestions for planners to consider (**Figure 5(a)** and **Figure 5(b)**). Firstly, public insights converge on key POI types, including Accommodation and Food Services, Real Estate and Rental, Retail Trade, and Arts, Entertainment, and Recreation. These findings underscore the need for planners to address density-related challenges linked to these specific POIs. Our analysis also presents specific insights for certain POI types. For example, locations featuring restaurants and drinking establishments, inegral to people's daily lives, require improved accessibility, which ensures easier access for individuals without the necessity of driving. This could be particularly important for restaurants located in densely urbanized regions (Area 3 in **Figure 4(a)**). In locales dominated by apartments, urban planners should prioritize community-oriented amenities like shopping centers, stores, recreational facilities, malls, and communal spaces. Meanwhile, areas with museums, art centers, and gyms need special attention to traffic and parking arrangements.

**Provide suggestions for equitable urban planning** – The PLS regression model demonstrates the socio-spatial roots of disparities in public sentiment toward urban density (**Table 4**). Broadly, minority communities and those with lower socioeconomic conditions tend to respond negatively to urban density, and denser land development exacerbates such negative sentiment. Interestingly, even when there is a mix of land development and denser transit services, it fails to significantly improve the public perception of community. This suggests that the negative effects induced by urban development outweigh the benefits gained from increased accessibility. This trend may stem from the disproportionate impacts of urbanization on different socioeconomic groups (Freeman & Braconi, 2004; Zuk et al., 2018). Affluent communities reap more benefits from land and facility development, such as improved accessibility, high-quality services, and numerous options and opportunities. Conversely, poor communities may bear the brunt of the negative impacts of urbanization, including parking challenges, traffic congestion, poor air and water quality, and rising service costs, leading to a deterioration of public attitudes toward urban density. Local governments and planners need to prioritize improving the built environment in socioeconomically disadvantaged communities by addressing key aspects such as accessibility and land development density. These strategies help ensure that the benefits brought by urban development are equitably distributed, fostering a sustainable and inclusive urban environment for all citizens.

## 5.3. Opportunities for future work

This study presents several avenues for future exploration. Firstly, many POIs may have very limited reviews. In our study, those POIs with less than 10 reviews were removed from the analysis to avoid the small sampling issue. However, this exclusion might overlook insights into urban density within areas or POI categories with sparse reviews. A potential future direction could involve amalgamating data from alternative platforms like Yelp, necessitating in-depth research for data fusion. Meanwhile, it's important to acknowledge that each online platform possesses distinct characteristics and trade-offs. Also, it could be interesting to examine the spatial patterns between the removed POIs and certain regions. Such an analysis could offer insights into why there are fewer density-related reviews posted from certain areas.

Secondly, the development of topic ontology and the establishment of training and testing datasets relied on a manual screening of the terms and reviews. This process involved the authors' interpretations of expressions related to urban density found in online reviews. However, it's conceivable that the developed ontology might not encompass all terms describing urban density, or there could be potential misclassifications within the samples. Subsequent research endeavors could explore expanding this ontology or datasets by involving more researchers to enhance the inclusivity and accuracy of the identified terms and training samples.

Thirdly, both the BERT-based models might encounter challenges in accurately identifying sentiment or classifying labels associated with reviews. For instance, the RoBERTa sentiment model was trained using tweet data (Loureiro et al., 2022), potentially not capturing the nuances specific to Google Maps reviews. Similarly, while our BERT sequence classification model was internally trained, its performance might be constrained due to limited training samples or time. A prospective avenue could involve integrating advanced large language models like GPT or LLaMa for text analysis.

Lastly, while leveraging crowdsourcing could mitigate biases inherent in selecting participants for an open-ended survey, it still introduced biases through the selection of individuals posting public comments. Prior studies indicated that younger and more educated demographics were prone to posting online reviews due to their familiarity with social media and online platforms (Mellon & Prosser, 2017; Wang et al., 2019). Furthermore, individuals with mostly positive or negative experiences were more inclined to post reviews, potentially leading to significant variance in feedback (Filieri, 2016).

There are several additional future avenues. One work could integrate survey data, such as Cheng et al. (2020) integrating household travel surveys with social media data to improve transport data quality, or leveraging information from diverse social media platforms like Twitter or Facebook. This could leverage a more comprehensive perspective from the residents by capturing a broader spectrum of opinions and experiences, potentially mitigating biases associated with the singular source of reviews. The other work could focus on the discussion between non-resident and resident accommodations. By examining the sentiment and experiences of individuals in both types of accommodations, we can gain insights into how living arrangements can shape people's behaviors and attitudes. For instance, non-resident accommodations such as hotels may prioritize convenience and novelty, whereas resident accommodations such as apartments may emphasize stability and community.

## 6. Conclusions

Our study illustrates the potential of passive crowdsourcing via online platforms to obtain public insights regarding urban density. Using Atlanta, Georgia as a case study, this study uses reviews for POIs from Google Maps and then employs multiple NLP techniques to process and analyze the textual information. Our study provides several important insights. Firstly, our study provides a geospatial map highlighting areas with distinct sentiment patterns – illustrating regions characterized by both positive and negative sentiments. Secondly, our study applies textual analysis to reveal consistent trends in attitudes across the four primary POI types. For example, negative sentiments from the public converge on issues related to parking and traffic management across these categories. Thirdly, our regression analysis uncovers intriguing associations: regions with more minorities and lower socioeconomic conditions exhibit more negative sentiments, while denser land development exacerbates such negativity. In conclusion, our study presents a scalable and transferable pipeline for sensing and measuring public attitudes, offering valuable insights in guiding livable, accessible, and equitable urban planning.

# Appendices

## A.1. "False" examples of Google Maps reviews

**Table A.1** shows the "False" examples of reviews. As explained in **Section 3.3**, these reviews contain typically descriptive terms in the topic ontology but are not informative of urban density.

Table A.1. Examples of "False" urban density reviews.

| Reviews | "False" urban density reviews |
|---|---|
| Food is just ok. Not good or delicious. There were soup bowls but no soup stall seen anyway near during lunch buffet. | There were soup bowls but no soup stall seen anyway near during lunch buffet. |
| Very over priced. Very slow service, automatic 20% gratituty. Staff very friendly. No social distancing. Every table was packed. Only 10% of the people had masks on [sic]. | Every table was packed. |
| Extremely useful combination of retail & food options to get your errands done in one spot. But the under ground parking area has been the target of some security issues and requires additional monitoring. | But the under ground parking area has been the target of some security issues and requires additional monitoring. |
| Love the atmosphere and vibe of this place. Restaurants and bars all throughout facility. Do not even have to be in the ball-park to enjoy the game. Can sit in a nearby restaurant in the complex and watch the game while you eat. Money was spent well building this facility. I would highly recommend this spot. | Can sit in a nearby restaurant in the complex and watch the game while you eat. |
| I've always enjoyed it, but recently most of the restaurants are closed due to low traffic volume from covid. | I've always enjoyed it, but recently most of the restaurants are closed due to low traffic volume from covid. |

## A.2. Grid search of hyperparameters for each classification model

**Table A.2** shows the parameter of the grid search for each classification model as developed in **Section 3.3**.

Table A.2. The grid search of hyperparameters for different ML classifiers.

| Model | Grid search range |
|---|---|
| TF-IDF + DT | Max depth: [5, 10, 20, 40, 60] <br> Min samples leaf: [1, 2, 4] |
| TF-IDF + RF | Num of estimators: [100, 200, 300, 400] <br> Max depth: [10, 20, 40, 80, 100] <br> Min samples leaf: [1, 2, 4] |
| TF-IDF + NB | Alpha (smooth parameter): [0.01, 0.05, 0.1, 0.2, 0.5, 1] |
| TF-IDF + SVM | Kernel: ['rbf', 'poly', 'linear'] <br> C (regularization parameter): [0.1, 0.5, 1, 2, 10] <br> Max iteration: [100, 200, 500, 1000, 1200] |
| TF-IDF + LR | Solver: ['sag', 'saga', 'lbfgs'] <br> C (inverse of regularization strength): [0.1, 0.5, 1, 2, 5, 10] <br> Max iteration: [10, 20, 50, 100, 200] |
| BERT | Num of epochs: [1, 2, 3, 4, 5] <br> Batch size: [16, 32, 64] |

**Note**: the selection of hyperparameters of each classifier is underlined "__" in the table.

## A.3. Partial least squares (PLS) regression

Key equations of PLS regression are presented below.

$$X = TP^T + E \qquad \text{Eq. A1}$$
$$Y = UQ^T + F \qquad \text{Eq. A2}$$
$$Y = XK^T + \Theta \qquad \text{Eq. A3}$$

where $X$ represents independent variables, including socio-demographics, transportation, and land use assumed to be correlated with the sentiment (refer to **Table 3** for details); $Y$ is the dependent variable, specifically the CBG-level average sentiment extracted from Google Reviews (refer to **Figure 3** for its distribution); Both $X$ and $Y$ are Z-score standardized, and the SIMPLS algorithm is employed to solve the equations (De Jong, 1993); $K$ is the coefficients; $T$ and $U$ are the orthogonal scores of $X$ and $Y$; $P$ and $Q$ are the orthogonal loadings; $E$ and $F$ represent error terms, assumed to be independently and identically distributed.

The decomposition process is iterative. In each iteration, data matrices ($X$ and $Y$) are deflated by subtracting the fitted components to generate new data matrices (i.e. the $E$ and $F$ become the new $X$ and $Y$ in the next iteration). The optimal number of iterations is determined by the cross-validated root mean squared error of prediction (RMSEP) (Mevik & Wehrens, 2007). In this study, the minimal RMSEP (0.240) was obtained at the first iteration. Hence, the number of components is selected as 1, achieving a percentage of variance explained of 85.94%.

## A.4. Lexical salience-valence analysis (LSVA)

The LSVA defines the salience and valence of a word as below (Taecharungroj & Mathayomchan, 2019),

$$salience|_{word_i} = log_{10}(N_{total})|_{word_i} \qquad \text{Eq. A4}$$
$$valence|_{word_i} = \frac{N(positive) - N(negative)}{N}|_{word_i} \qquad \text{Eq. A5}$$

where
  $N_{total}$ represents the total number of reviews that $word_i$ appears
  $N(positive)$ denotes the number of positive reviews that $word_i$ appears
  $N(negative)$ denotes the number of negative reviews that $word_i$ appears

The salience of a word is computed by the logarithm with a base 10 function of the frequency of each term. The valence of a word is computed as $N(positive) - N(negative)$ divided by its total count $N_{total}$, which measures how positive a particular word is in a corpus. Reviews that contain words with highly positive valence are more likely to be positive reviews than those with negative words.